\documentclass[showpacs,aps,showkeys]{revtex4}

\usepackage{graphicx}

\usepackage[usenames]{color}
\usepackage{cancel}
\usepackage{hyperref}
\usepackage[normalem]{ulem}
\usepackage{amsfonts}
\usepackage{amsmath}
\usepackage{graphicx}
\usepackage{epstopdf}
\usepackage{float}
\usepackage{enumerate}
\usepackage{graphics}

\begin{document}

\title{The solution space structure of planted constraint satisfaction problems with growing domains}

\author{Wei Xu$^1$ Zhe Zhang$^2$}
\email[]{s20190769@xs.ustb.edu.cn}
\affiliation{
$^{1}$$^{2}$School of Mathematics and Physics, University of Science and Technology Beijing, Beijing 100083, China\\
}

\date{\today}

\begin{abstract}
Planting a solution into the random RB model, which is a prototype of random constraint satisfaction problem (CSP) with growing domains, can generate very hard satisfiable CSP benchmarks. We study the solution space structure of the planted RB model. With constraint density growing, we find that this model goes through four phase transitions. In the replica symmetric phase, what we call the independent phase transition occurs, after which the planted cluster (cluster containing the planted solution) is separated from the giant cluster. Then the solutions except that in the planted cluster go through the same clustering phase transition and the same satisfiability phase transition as the random RB model. The planted cluster goes through the isolated phase transition, after which the planted cluster contains only one solution. This phase diagram provides strong evidence that this model can generate very hard satisfiable CSP benchmarks. For over constraint instances (where the constraint density is very large), we find that the configuration space has only a single energy valley, which makes the instances tractable. Experiments using Belief Propagation confirm the locations of the clustering, satisfiability (by configurations outside the planted cluster), and isolated phase transition points.
\end{abstract}
\keywords{constraint satisfaction problem, solution space structure, phase transition, problem hardness, belief propagation.}
\pacs{89.75.Fb, 02.50.-r, 64.70.P-, 89.20.Ff}
\maketitle

\section{introduction}\label{sec:pair}
The constraint satisfaction problems (CSPs) play a significant role in computer science, statistical physics and mathematics. The average computational complexity of CSPs has been studied from the point of view of spin glass theory \cite{oxford}, and it is related to the phase diagram. A planted CSP is also an inference problem , and the computational feasibility of inference problem is also related to its phase diagram \cite{inference}.

Timetabling, hardware configuration, factory scheduling, floorplanning and many other tasks, can be solved by a unified process called ``constraint programming'': firstly those tasks are modeled as CSPs and secondly the CPSs are solved by computer algorithms \cite{handbook}. For a better understanding of this unified process, it is necessary to generate CSP instances randomly, and for this propose the classical models A, B, C and D are proposed \cite{Gent}\cite{Smith}. But Achlioptas atc. \cite{ach97} found that those models suffer trivial asymptotic insolubility: asymptotically almost all instances they generate have no solutions. To overcome this deficiency, two main approaches are applied: one is incorporating some structures (e.g. arc-consistent), another is controlling the way parameters change as the problem size increases\cite{Lecoutre}. The random RB (revised B) model \cite{xu2000} with growing domains follows the second approach.

To benchmark algorithms, not only random CSP models are required, but also models that generate satisfiable instances. If an incomplete algorithm does not find a solution for a satisfiable instance, there is no doubt that the algorithm fails. Many efforts have been done to generate satisfiable instances, including that from the physical point of views \cite{bar2002,jia2004,locked}. Planting is a simple way to hide a solution in CSPs: a solution $S$ to be hidden is chosen in advance, then only instances that have $S$ as a solution are generated. In cryptographic application, hard planted CSP instances serve as one-way functions. Recovering the planted solution is also in the category of inference problem, which have gotten intense attentions \cite{inference}.

Benchmarks based on the planted RB model \cite{ke} have been used in various kinds of algorithm competitions (e.g. CSP, SAT and MaxSAT), and the results confirmed the intrinsic hardness of these benchmarks. Based on this model, Ke Xu \cite{web} proposed an instance ``frb100-40'' with 100 variables in 2005 and challenged that the instance can not been solved on a PC in less than 24 hours in 20 years. The challenge is still continuing after several researchers tried \cite{cai,rosin}.

CSPs are related to the spin glass theory naturally. A CSP instance contains a set of discrete variables and a collection of constraints. A constraint restricts the joint values of some variables, so a constraint acts as an interaction among some variables. A Gibbs measure can be defined, where the energy of a configuration (an assignment of all variables) represents the number of constraints that the configuration violates.
Solutions are zero energy configurations. At zero temperature, the partition function is the number of solutions. A CSP model defines a distribution to a set of CSP instances.

From the point of view of statistical physics, the average computation hardness is related to the solution space structure. The phase diagrams of many CSP models have been studied  \cite{oxford}. Usually a CSP model goes through the replica symmetric (RS) phase, where almost all solutions belong to a giant cluster; the dynamic one step replica symmetry breaking (d1RSB) phase (or so called the clustering phase), where solutions shatter into exponentially many clusters; the static one step replica symmetry breaking (s1RSB) phase, etc.

In this paper, the first moment method is frequently used. Our results on solution space structure are established on typical instances, i.e. they hold ``with high probability (w.h.p.)'', which means ``with probability $1-o(1)$ with the problem size $n\rightarrow \infty$''. As we always study properties on typical instances, ``w.h.p.'' is omitted if there is no ambiguity.

In this paper, we find that the planted RB model goes through the independent, clustering, satisfiability (by configurations outside the
planted cluster), and isolated phase transitions. The main result is in FIG. \ref{d2}. The rest of the paper is organized as follows: we give definitions of the random model and the planted RB model in Sec. \ref{sec:model}, then
study the independent phase transition and the satisfiability (by configurations outside the
planted cluster) phase transition in Sec. \ref{sec:id}, study  the clustering phase transition in Sec. \ref{sec:ds}, study  the isolated phase transition in Sec. \ref{sec:is}, then draw the phase diagram in Sec. \ref{sec:pd}. In Sec. \ref{sec:over}, we study the energy valleys of over constraint instances. In Sec. \ref{sec:ex}, we do some experiments and study the fixed points of belief propagation equations and the solutions found by BP guided algorithms.

\section{Definitions of the random and planted RB model}\label{sec:model}

An instance of the RB model is comprised of $n$ variables and $t$ constraints. All the $n$ variables take values from a domain $D=\{1,2, ...,d\}$, where $d=n^\alpha, \alpha>0$. Each of the $t$ constraints involves $k$ variables and restricts the tuples of values of the $k$ variables. For a constraint, the set of compatible tuples of values is a subset of $D^k$. A tuple of values of all the $n$ variables is called an assignment (configuration). An assignment is a solution if it satisfies all the $t$ constraints.
We give the definitions of the RB model and the planted RB model by their steps to generate instances. A random RB instance is generated by the following two steps \cite{xu2000}:
\begin{enumerate}
\item Select with repetition $t= rn\ln n$ random constraints. Each constraint is formed by randomly
selecting without repetition $k$ of $n$ variables.

\item For each constraint, randomly select without repetition $(1-p) d^k$ ($0 < p < 1$ measures
the tightness of the constraint) compatible tuples of values.
\end{enumerate}
A planted RB instance is generated by the following three steps:
\begin{enumerate}
\item Choose a random assignment $S$ as the planted solution.
\item This step is as same as step 1 of random RB model.
\item For each constraint, randomly select without repetition $(1-p)d^k$ compatible tuples of values, where the tuple of values of the $k$ variable in $S$ must be included.
\end{enumerate}

For the random RB model, when $p<1-\frac{1}{k}$ and $\alpha>1/k$, the satisfiability transition occurs at $r_s$ \cite{xu2000}:
\begin{equation*}
r_s=-\frac{\alpha}{\ln(1-p)}.
\end{equation*}
In the following, we let $r=r_0r_s$, then the satisfiability transition happens at $r_0=1$. In the following, we only consider models under the condition: $p<1-\frac{1}{k}$ and $\alpha>1/k$.

We should list some definitions which follow \cite{pair1}. The (Hamming) distance between two assignments (configurations) $A$ and $B$ is the number of variables where $A$ and $B$ take different values. $A$ and $B$ are connected if and only if the distance between them is 1. Cluster is connected component of solutions. Cluster-region is set of clusters. Diameter of a cluster-region is the biggest distance between two solutions in the region.

In Sec. \ref{sec:id} we find that after the independent phase transition, a set of solutions centered on $S$ is separated from other solutions. This set includes solutions being at distance smaller than $\epsilon n$ from $S$, where $\epsilon$ is arbitrary small positive constant. Because this set of solutions is very small, we might as well call the set a cluster: the planted cluster.

\section{the independent phase transition and the transition at $r_s$}\label{sec:id}

In this section we will show that, in the replica symmetric phase, what we call the independent phase transition occurs, before which the planted solution belongs to the giant cluster, and after which the planted cluster (cluster containing
the planted solution) is far away from the giant cluster.

Let $E(Y(x))$ be the average number of solutions being at distance $xn$ from the planted solution $S$, then by the definition of the planted RB model,
\begin{align}
E(Y(x))=\left(\begin{array}{c}n\\xn,n-xn\end{array}\right)(d-1)^{xn}\left(\hat{p}\right)^{rn\ln n}, \label{yx}
\end{align}
where
\begin{align}
\hat{p}=\frac{\left(\begin{array}{c}n-xn\\k,n-xn-k\end{array}\right)}{\left(\begin{array}{c}n\\k,n-k\end{array}\right)}+\frac{(1-p)d^k-1}{d^k-1}\left(1-\frac{\left(\begin{array}{c}n-xn\\k,n-xn-k\end{array}\right)}{\left(\begin{array}{c}n\\k,n-k\end{array}\right)}\right),\label{hatp}
\end{align}
and $\left(\begin{array}{c}n\\m,n-m\end{array}\right)$ repesents combination formula $\frac{n!}{m!(n-m)!}$.
Let $$h(x)=\lim_{n\rightarrow\infty}\frac{\ln E(Y(x))}{\alpha n\ln n}.$$
For constant $0<x<1$, with $n\rightarrow \infty$, we have
\begin{align*}
&\frac{(1-p)d^k-1}{d^k-1}\rightarrow 1-p,\\
&\frac{\left(\begin{array}{c}n-xn\\k,n-xn-k\end{array}\right)}{\left(\begin{array}{c}n\\k,n-k\end{array}\right)}\rightarrow (1-x)^k,\\
&\frac{\ln{\left(\begin{array}{c}n\\xn,n-xn\end{array}\right)}}{n}\rightarrow -\ln(x^x(1-x)^{1-x}),
\end{align*}
where the last one is from the Stirling formula.
Then we have
\begin{align*}
h(x)& = \lim_{n\rightarrow\infty}\left[\frac{-\ln(x^x(1-x)^{1-x})}{\alpha\ln n}+\frac{x\ln (d-1)}{\alpha\ln n}+\frac{r_0}{-\ln (1-p)}\ln \left(1-p+p(1-x)^k\right)\right]\\
&= x+\frac{r_0}{-\ln (1-p)}\ln \left(1-p+p(1-x)^k\right).
\end{align*}
\begin{figure}[htbp]
  \centering
  \includegraphics[width=0.6\columnwidth]{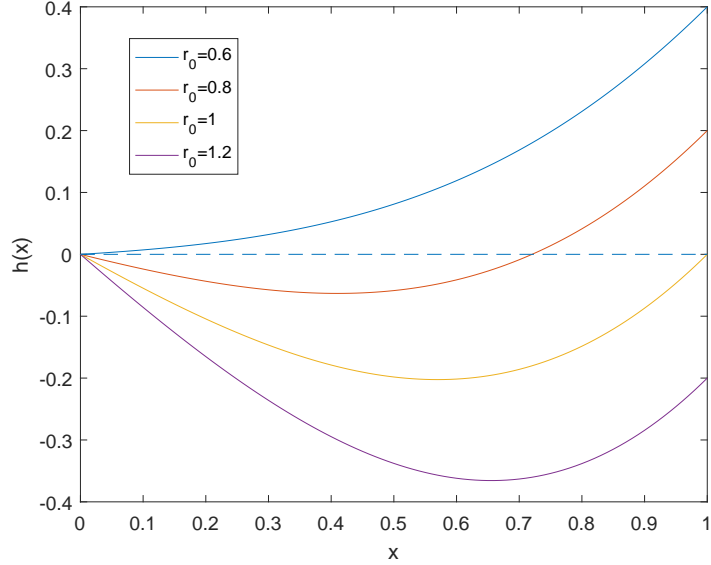}
  \caption{Function images of $h(x)$ when $k= 2$, $p=0.4$ and $r=0.6r_s$, $r=0.8r_s$, $r=1r_s$, $r=1.2r_s$ from top to bottom.}\label{hx}
\end{figure}
In FIG. \ref{hx} we draw the function images of $h(x)$ for $k= 2$, $p=0.4$ and different $r$s. We calculate the first and second order derivatives of $h(x)$:
\begin{align*}
\frac{dh(x)}{dx}=1+\frac{r_0pk(1-x)^{k-1}}{\ln(1-p)\left(1-p+p(1-x)^k\right)},
\end{align*}
\begin{align*}
\frac{d^2h(x)}{dx^2}=\frac{r_0pk(1-x)^{k-2}[(k-1)(1-p)-p(1-x)^k]}{\left(-\ln(1-p)\right)\left(1-p+p(1-x)^k\right)^2}.
\end{align*}
We only consider the model under condition $k\geq \frac{1}{1-p}$, which leads to $(k-1)(1-p)-p\geq 0$ and $(k-1)(1-p)-p(1-x)^k>0$ ($0<x<1$), then $h(x)$ is concave: $$\frac{d^2h(x)}{dx^2}>0.$$

We study the following three situations:

1. $\frac{dh}{dx}|_{x=0}>0$. It is easy to verify that $h(0)=0$. If $\frac{dh}{dx}|_{x=0}>0$, then $h(x)>0$ for all constant $0<x<1$.

2. $\frac{dh}{dx}|_{x=0}<0$ and $r<r_s$ ($r_0<1$). We have $h(0)=0$, $h(1)=1-r_0>0$, $\frac{dh}{dx}|_{x=0}<0$ and the concavity of $h(x)$. The function image of $h(x)$ will be like the $r=0.8r_s$ line in FIG. \ref{hx}. We have $f(x)<0$ for $\epsilon\leq x <a$, where $\epsilon$ is arbitrary small positive constant, $a$ is the positive solution of $h(x)=0$. Let $b$ be a little smaller than $a$, then $f(x)<h_0$ for $\epsilon\leq x \leq b$, where $h_0=max\{h(\epsilon),h(b)\}<0$.
$$\lim_{n\rightarrow \infty}E\left(\sum_{\epsilon n\leq xn\leq bn}Y(x)\right)=\lim_{n\rightarrow \infty}\sum_{\epsilon n\leq xn\leq bn}E\left(Y(x)\right)\leq \lim_{n\rightarrow \infty}ne^{h_0\alpha n\ln n}=0.$$
By the first moment method we have w.h.p. $\sum_{\epsilon n\leq xn\leq bn}{Y(x)}=0$.

3. $r>r_s$ ($r_0>1$). We have $h(0)=0$, $h(1)=1-r_0<0$, and the concavity of $h(x)$. For arbitrary small positive constant $\epsilon$, we have $f(x)<0$ for $\epsilon\leq x \leq1$. Let $h_1=max\{h(\epsilon),h(1)\}$, we have $h_1<0$ and
$$\lim_{n\rightarrow \infty}E\left(\sum_{\epsilon n\leq xn\leq n}Y(x)\right)=\lim_{n\rightarrow \infty}\sum_{\epsilon n\leq xn\leq n}E\left(Y(x)\right)\leq \lim_{n\rightarrow \infty}ne^{h_1\alpha n\ln n}=0.$$
By the first moment method we have that w.h.p. $\sum_{\epsilon n\leq xn\leq n}Y(x)=0$.

\textbf{The independent phase transition.} From situation 2, we find for typical instances, the solutions being at distance $xn (\epsilon n\leq xn\leq bn)$ from the planted solution $S$ do not exist, where $\epsilon$ is arbitrary small positive constant. The solution space can be split into two parts: the first part centered on $S$ includes solutions being at distance smaller than $\epsilon n$ from $S$; the second part includes other solutions. Because $\epsilon$ can take arbitrarily small positive value, the first part is very small, then we might as well call the first part a cluster: the planted cluster.

Equation $\frac{dh}{dx}|_{x=0}=0$ defines the independent transition point, denoted by $r_{id}$. When $r<r_{id}$, the planted solution is in the giant cluster (actually we do not prove this mathematically in this paper). When $r>r_{id}$, the planted cluster is separated from the other solutions. Solving equation $\frac{dh}{dx}|_{x=0}=0$, we have $r_0=\frac{-\ln(1-p)}{kp}$, so $$r_{id}=r_0r_s=\frac{-\ln(1-p)}{kp}r_s=\frac{\alpha}{kp}.$$

It is easy to verify that $h(x)=g(x)/\alpha-(1-r_0)$, where $g(x)$ is defined in the following equations (\ref{eq:gx}) and (\ref{eq:f1}). By this relation, it is found that, with $r_0$ growing, ``$\frac{dh}{dx}|_{x=0}=0$'' occurs before ``$g(x)=0$ for some constant $x$''. This means that the independent phase transition is before the clustering phase transition.

\textbf{The satisfiability (by configurations outside the planted cluster) phase transition.} From situation 3, we find for typical instances, no solutions are at a distance bigger than $\epsilon n$ from $S$, where $\epsilon$ is arbitrary small positive constant. It is to say $r_s$ is a transition point, after which (for typical instances) no solutions exist except that in the planted cluster. Before $r_s$, the model is in the clustering phase, and there are exponentially many clusters, seeing Sec. \ref{sec:ds}.

\section{the clustering phase transition}\label{sec:ds}
In this section, we will show that the planted RB model has the same clustering phase as the random RB model.
The method that we will apply to study the clustering phase has no differences to what was used on the random RB model in \cite{xuwei}. The method focuses on the number of solution-pairs at certain distances. If solution-pairs at distance $xn, an<xn<bn$ do not exist, then by the method the solution space shatters into cluster-regions. This method has been applied on random graph coloring, random k-SAT, random hypergraph 2-coloring, random RB model, and random d-k-CSP model \cite{pair1,pair2,pair3,xuwei,xuwei2}.

\begin{figure}[htbp]
  \centering
  \includegraphics[width=0.6\columnwidth]{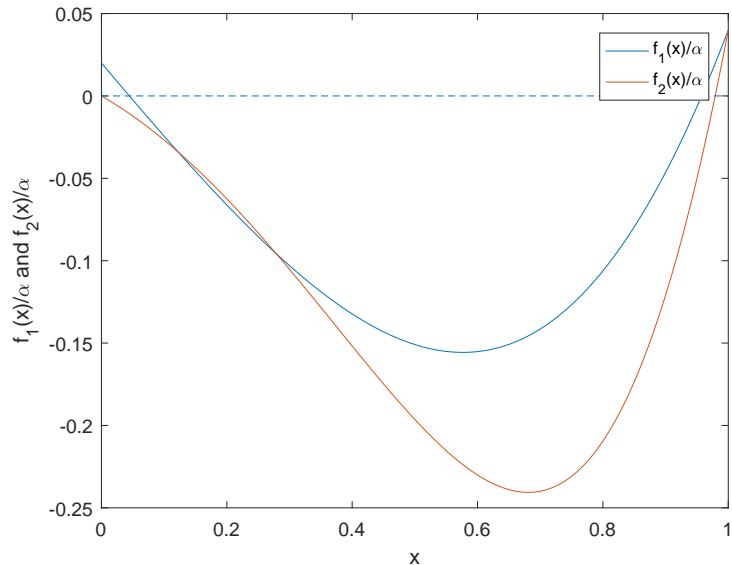}
  \caption{Function images of $f_1(x)/\alpha$ and $f_2(x)/\alpha$ when $k= 2$, $p=0.47$ and $r=0.98r_s$.}\label{f1f2}
\end{figure}

Denote the expectation of the number of solutions pairs at distance $xn$ by $\mathbb{E}(Z(x))$, and define its normalised version as
\begin{eqnarray}\label{eq:fx0}
f(x)=\lim_{n \rightarrow \infty}{\ln{\left(\mathbb{E}(Z(x))\right)}/(n\ln n)}.
\end{eqnarray}
As long as $f(x)<0$, we can conclude that the typical instances do not have solution-pairs at distance $xn$.
In appendix, we show that
\begin{eqnarray}\label{eq:fx1}
f(x)=max(f_1(x), f_2(x)),
\end{eqnarray}
where
\begin{eqnarray}\label{eq:f1}
f_1(x)=\alpha (1+x)+r\ln[(1-p)^2+(1-p)p(1-x)^k]
\end{eqnarray}
\begin{eqnarray}\label{eq:f2}
f_2(x)=2\alpha x+r\ln[(1-p)^2+(2p-p^2)(1-x)^k]
\end{eqnarray}
We take $k= 2$, $p=0.47$ and $r=0.98r_s$ for an example, and draw the function images of $f_1(x)/\alpha$ and $f_2(x)/\alpha$ in FIG. \ref{f1f2}. In the figure we find that when $x\in(0.038, 0.958)$ we have $f(x)<0$, which means that the typical instances do not have solution-pairs at distance between $0.038n$ and $0.958n$. The solution space can be divided into cluster-regions, where the diameter of a cluster-region is at most $0.038n$ and the distance between two cluster-regions is at least $0.958n$ \cite{xuwei}.

\begin{figure}[htbp]
  \centering
  \includegraphics[width=0.6\columnwidth]{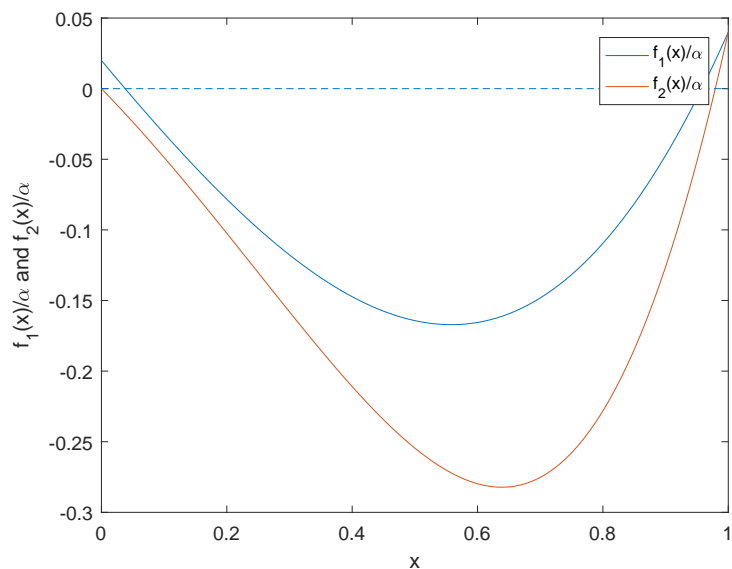}
  \caption{Function images of $f_1(x)/\alpha$ and $f_2(x)/\alpha$ when $k= 2$, $p=0.4$ and $r=0.98r_s$.}\label{dui2}
\end{figure}

Let $g(x)$ be the counterpart of $f(x)$ in the random RB model, i.e. the logarithm of the expectation of the number of solutions pairs at distance $xn$ divided by $n\ln n$ in the random RB model, then the equation (5) in \cite{xuwei} shows that
\begin{eqnarray}\label{eq:gx}
g(x)=f_1(x).
\end{eqnarray}
We draw a lot of function images of $f_1(x)$ and $f_2(x)$ for different parameters $k\geq 2$, $0<p<1-1/k$ and $0< r< r_s$. Numerical calculation shows that $f_1(x)$ and $f(x)$ have the same positive part, i.e. $f_1^+(x)=f^+(x)$, then
\begin{eqnarray}\label{eq:g+}
g^+(x)=f^+(x).
\end{eqnarray}
In most cases, $f_1(x)>f_2(x)$ for $0\leq x\leq 1$, as shown in FIG. \ref{dui2}. In a few cases, $f_1(x)<f_2(x)$ for some $x$ but for those $x$ we have $f_1(x)<f_2(x)<0$, as shown in FIG. \ref{f1f2}. In all the cases we have that $f_1^+(x)>f_2^+(x)$ then $f_1^+(x)=f^+(x)$ and $g^+(x)=f^+(x)$.

Equation (\ref{eq:g+}) means that $f(x)<0$ and $g(x)<0$ hold on the same interval of $x$. By the method splitting the solution space, we find that the planted RB model has the same clustering phase as the random model. The clustering region is $(r_d, r_s)$, where $r_d$ is the smallest value of $r$ for which $g(x)=0$
has at least one solution in $x\in[0, 1]$.

The number of solutions in a cluster-region is limited by the diameter of the cluster-region. In \cite{xuwei}, it is found that the typical instance of random RB model have a lot solutions (when $r<r_s$), and one cluster-region contains only an exponentially small proportion of them. For the planted model, the diameter of cluster-region remains unchanged. But instances of the planted model is generated with probability proportional to the number of its solutions, so they tend to have more solutions then that generated from the random model. We conclude that for the planted model, the number of cluster-regions is exponential and one cluster-region contains an exponentially small proportion of solutions.

\section{the isolated phase transition}\label{sec:is}
In this section we show that with $r$ growing, the planted cluster goes through a transition, after which the planted solution has no solutions as neighbors. We call it the isolated transition.

The average number of solutions being at distance $xn$ from the planted solution $S$, $E(Y(x))$, was given in equation (\ref{yx}). For any $xn=1,2,...,n$, we have
$$\frac{C_{n-xn}^{k}}{C_{n}^{k}}=\frac{(n-xn)...(n-xn-k+1)}{n...(n-k+1)}<(1-x)^k.$$
Combining
$$\frac{(1-p)d^k-1}{d^k-1}<1-p,$$
for any $xn=1,2,...,n$, we have
$$E(Y(x))\leq n^{xn}d^{xn}\left(1-p+p(1-x)^k\right)^{rn\ln n},$$
and
\begin{align*}
\frac{\ln E(Y(x))}{xn\ln n}&\leq 1+\alpha+r\frac{\ln \left(1-p+p(1-x)^k\right)}{x}.
\end{align*}
Because
\begin{align*}
\lim_{x\rightarrow 0}\frac{\ln \left(1-p+p(1-x)^k\right)}{x}=-kp,
\end{align*}
we have
\begin{align*}
&\lim_{x\rightarrow 0}\frac{\ln E(Y(x))}{xn\ln n}\leq 1+\alpha-rkp.
\end{align*}
Let $\omega=1+\alpha-rkp$. When $\omega<0$, there is a constant number $\epsilon$, such that for $0<x\leq\epsilon$ we have $$\frac{\ln E(Y(x))}{x n\ln n}<\frac{1}{2}\omega.$$
Then we have $$E\left(\sum_{1\leq xn\leq\epsilon n}Y(x)\right)=\sum_{1\leq xn\leq\epsilon n}E\left(Y(x)\right)\leq \sum_{x n=1,2,...}E\left(Y(x)\right)<n^{1*\frac{1}{2}\omega}+n^{2*\frac{1}{2}\omega}+...\leq\frac{n^{\frac{1}{2}\omega}}{1-n^{\frac{1}{2}\omega}}\xrightarrow{n\rightarrow \infty} 0. $$

By the first moment method, as long as $\omega<0$ (i.e. $r>\frac{1+\alpha}{kp}$), we have w.h.p. $\sum_{1\leq xn\leq\epsilon n}Y(x)=0$. For typical instances, there are no solutions being at a distance smaller than $\epsilon n$ ($\epsilon$ is a positive constant number depending on $\omega$) from the planted solution, except the planted solution itself. We say $$r_{is}=\frac{1+\alpha}{kp}$$ is the isolated transition point, above which the planted cluster shrinks into a point (the planted solution).

When $r>r_s$ and $r>\frac{1+\alpha}{kp}$, for typical instances, the planted solution is the only solution.
\section{phase diagram of the planted RB model}\label{sec:pd}
By the analyses in the above three sections, we find four phase transitions of the planted RB model: the independent transition at $r_{id}=\frac{\alpha}{kp}$, the clustering transition at $r_{d}$, the satisfiability (by configurations outside the
planted cluster) transition at $r_s=-\frac{\alpha}{\ln(1-p)}$, the isolated transition at $r_{is}=\frac{1+\alpha}{kp}$. $r_d$ is the smallest value of $r$ for which $g(x)=0$
has at least one solution in $x\in[0, 1]$, where $g(x)$ is defined in equations (\ref{eq:gx}) and (\ref{eq:f1}).

\begin{figure}[htbp]
  \centering
  \includegraphics[width=0.8\columnwidth]{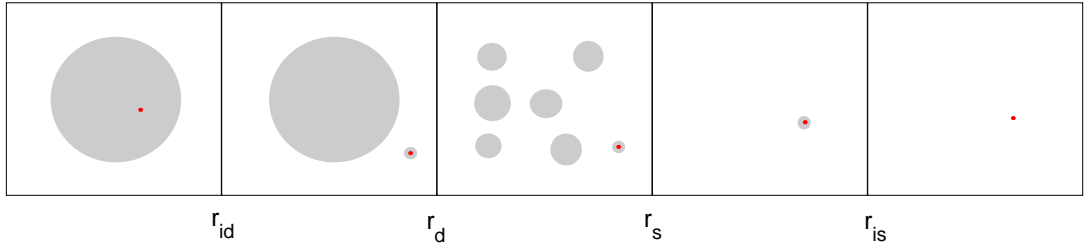}
  \caption{Phase diagram of the planted RB model in the $r_{is}>r_s$ case. The planted RB model goes through the independent transition at $r_{id}=\frac{\alpha}{kp}$, the clustering transition at $r_{d}$, the satisfiability (by configurations outside the planted cluster) transition at $r_s=-\frac{\alpha}{\ln(1-p)}$ and the isolated transition at $r_{is}=\frac{1+\alpha}{kp}$. $r_d$ is the smallest value of $r$ for which $g(x)=0$
has at least a solution in $x\in[0, 1]$, where $g(x)$ is defined in equation (\ref{eq:gx}). The plant solution is the red point. Diameter of the planted cluster is smaller than $\epsilon n$ for any constant $\epsilon>0$. }\label{d2}
\end{figure}

The phase diagram is drawn in FIG. \ref{d2} in the $r_{is}>r_s$ case.
Below $r=r_{id}$, most of the solutions, including the planted solution, belongs to a giant cluster. Above $r=r_{id}$, the planted cluster is separated from the giant cluster, and the distance between them is linear to $n$. Diameter of the planted cluster is smaller than $\epsilon n$, where $\epsilon$ is arbitrary small positive constant. From then on, the planted cluster and the other solutions have different transitions. Above $r=r_d$, solutions shatter into exponentially many clusters, and each cluster contains only an
exponentially small proportion of solutions. Above $r=r_s$, no solutions exist except that in the planted cluster. Above $r=r_{is}$, the planted cluster has only the planted solution left.

\begin{figure}[bp]
  \centering
  \includegraphics[width=0.4\columnwidth]{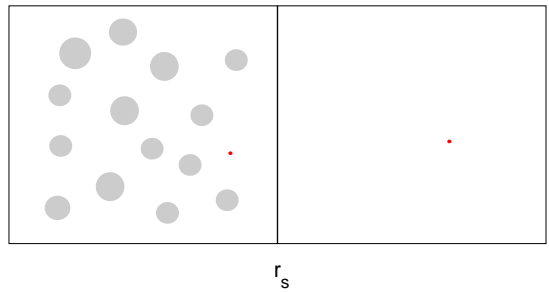}
  \caption{Phase diagram of the planted RB model around $r_s$ in the $r_{is}<r_s$ case. The planted solution (the red point) is isolated around $r_s$.}\label{d3}
\end{figure}

$r_{is}<r_d$, $r_d<r_{is}<r_s$, and $r_{is}>r_s$ are all possible, depending on the parameters $\alpha, p, k$.
In FIG. \ref{d3}, we show the phase diagram around $r=r_s$ in the $r_{is}<r_s$ case. Below $r=r_s$, there are many cluster all around the solution space. Above $r=r_s$ only the planted solution remains.

The phase diagram provides strong evidence that this model can generate very hard satisfiable CSP benchmarks. There are three reasons: 1. Below $r=r_s$, the planting is quiet, and the planted RB model has the same clustering phase as the random model, so the planting does not reduce the problem hardness. 2. Above $r=r_s$, solutions disappear except that in the planted cluster. Sudden disappearance of most of the solutions will make the problem hard. 3. Above $r=r_s$, the planted cluster is small, so the planted solution tends not to be exposed by its neighbors.

We propose a condition under which the problem should be harder (at $r=r_s$). The condition is $r_{is}<r_s$ or equally $$\frac{1+\alpha}{kp}<-\frac{\alpha}{\ln(1-p)}.$$ Under this condition, the phase diagram around $r=r_s$ has been shown in FIG. \ref{d3}. The planted solution is isolated around $r_s$, so it should be more difficult to be found.

\section{energy valley of over constraint instances}\label{sec:over}

\begin{figure}[htbp]
  \centering
  \includegraphics[width=0.6\columnwidth]{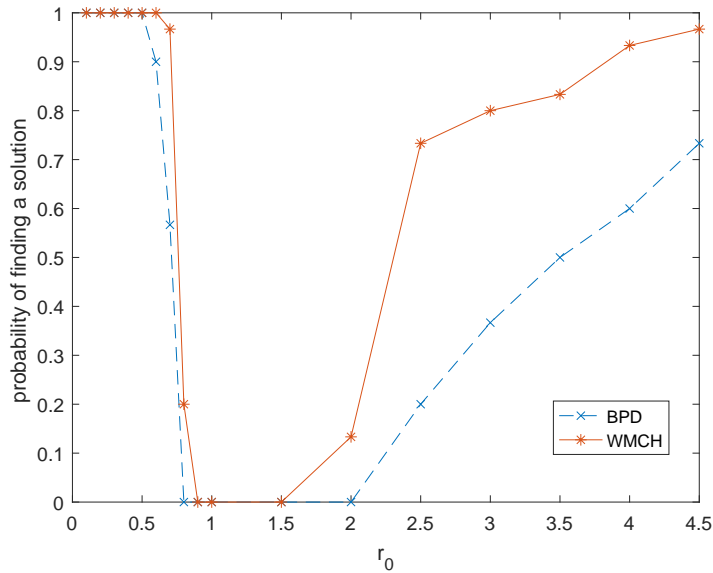}
  \caption{Belief propagation decimation (BPD) algorithm and the WMCH algorithm on the planted model with $n=100, k=2, \alpha=0.6, p=0.4, r=r_0r_s$.}\label{nanyinan}
\end{figure}

We perform the BPD (Belief propagation decimation) algorithm and the WMCH (min conflicts heuristic with random walk, referring to Chapter 5 of \cite{handbook}) algorithm on the planted RB model with $n=100, k=2, \alpha=0.6, p=0.4$. From FIG. \ref{nanyinan}, we find an easy/hard/easy pattern, and that when $r\gg r_s$ the instances are tractable. This pattern has been studied on other models \cite{hiding}\cite{why}.

We study the energy of configuration, which is the number of constraints that the configuration violates. For the planted RB model, let $E^{x,\lambda}(X)$ be the average number of configurations with energy $\lambda t$ and at distance $x n$ from $s$, then
\begin{align*}
E^{x,\lambda}(X)=\left(\begin{array}{c}n\\xn,n-xn\end{array}\right)(d-1)^{nx}\left(\begin{array}{c}t\\ \lambda t,t-\lambda t\end{array}\right)\hat{p}^{(1-\lambda) t}(1-\hat{p})^{\lambda t},
\end{align*}
where $\hat{p}$ is defined in (\ref{hatp}) and $t=rn\ln n$.
Let $$f(x,\lambda)=\lim_{n\rightarrow \infty}{\frac{\ln E^{x,\lambda}(X)}{n\ln d}},$$ we can simplify the values of $f(x,\lambda)$ for constant number $x$:
\begin{align*}
f(x,\lambda)=x+\phi(\lambda)r_0,
\end{align*}
where
\begin{align*}
\phi(\lambda)=\frac{1}{\ln (1-p)}\ln(\lambda^\lambda(1-\lambda)^{1-\lambda})-(1-\lambda)\frac{\ln(1-p+p(1-x)^{k})}{\ln (1-p)} -\lambda\frac{\ln(p-p(1-x)^{k})}{\ln (1-p)}.
\end{align*}
Calculating the first and second order derivatives, we have
$$\frac{\partial \phi(\lambda)}{\partial \lambda}=\frac{1}{\ln (1-p)}[\ln \lambda-\ln(1-\lambda)]+\frac{\ln(1-p+p(1-x)^{k})}{\ln (1-p)} -\frac{\ln(p-p(1-x)^{k})}{\ln (1-p)},$$
$$\frac{\partial^2 \phi(\lambda)}{\partial \lambda^2}=\frac{1}{\ln p}(\frac{1}{\lambda}+\frac{1}{1-\lambda})< 0.$$
We find $\phi(\lambda)$ is convex, and achieves its maximum 0 at $\lambda=p-p(1-x)^{k}$.

\begin{figure}[htbp]
  \centering
  \includegraphics[width=0.4\columnwidth]{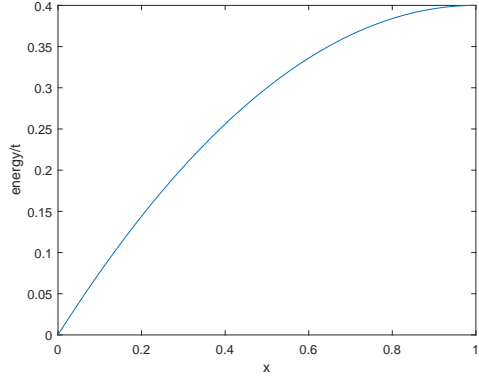}
  \caption{The energy divided by $t$ of configurations at distance $xn$ to the planted solution, when $k=2, p=0.4$.}\label{energy}
\end{figure}

For all constant number $0\leq x \leq 1$ there are $\lambda^{*}=p-p(1-x)^{k}$ and $r^{*}=\frac{-x}{\phi(\lambda)}$, when $\lambda\neq \lambda^*$ and $r>r^*r_s$, we have $f(x,\lambda)<0$. By the first moment method, $f(x,\lambda)<0$ means that (w.h.p.) configurations with energy $\lambda t$ and at distance $x n$ from $S$ do not exist.
So when $r$ is big enough, the configurations being at distance $xn$ from the planted solution $S$ have energy  $$\lambda^{*}t=(p-p(1-x)^{k})t.$$  From FIG. \ref{energy}, we can see that the configuration space has only one valley, and that is the reason that local search algorithm can find a solution fast.

\section{experiments using belief propagation}\label{sec:ex}
We take $n=50, k=2, \alpha=0.8, p=0.4$ for an example, and study its solution space structure by experiments. For the parameters $k=2, \alpha=0.8, p=0.4$, we can calculate the transition points: the independent transition point $r_{id}=\frac{-\ln(1-p)}{pk}r_s=0.6385r_s$, the clustering transition point $r_{d}=0.8815r_s$ (referring to FIG. 1 of \cite{xuwei}), the satisfiability (by configurations outside the
planted cluster) transition point $r_s$, and the isolated transition point $r_{is}=\frac{1+\alpha}{kp}=1.4367r_s$.

We study the belief propagation (BP) equations on this example. The BP equations \cite{oxford} are
\begin{align*}
\nu_{i\rightarrow a}(x_i)=\frac{1}{Z_{i\rightarrow a}}\prod_{b\in \partial i\setminus a} \widehat{\nu}_{b\rightarrow i}(x_i),\\
\widehat{\nu}_{a\rightarrow i}(x_i)=\frac{1}{Z_{a\rightarrow i}}\sum_{\underline{x}_{\partial a\setminus i}}\psi_a(\underline{x}_{\partial a})\prod_{k\in \partial a\setminus i} \nu_{k\rightarrow a}(x_k),
\end{align*}
where $\nu_{i\rightarrow a}$ is the message from variable $i$ to constraint $a$, $\widehat{\nu}_{a\rightarrow i}$ is the message from constraint $a$ to variable $i$. $\psi_a$ is equal to 1 if constraint $a$ is satisfied by $\underline{x}_{\partial a}$, and is equal to 0 otherwise. $\partial a$ denotes the set of variables connected to constraint $a$, and $\partial a\setminus i$ is the set removing $i$.  $Z_{i\rightarrow a}$ and $Z_{a\rightarrow i}$ are normalization constants, and here if they equal to 0, we set the messages equal to $1/d$ for all $x_i=1,...,d$.

Solving BP equations by iteration, if the iteration converges, the messages $\nu_{i\rightarrow a}$, $\widehat{\nu}_{a\rightarrow i}$ arrive at a fixed point. For a fixed point, marginal probabilities of variable $i$ are
\begin{align*}
\nu_{i}(x_i)\cong\prod_{b\in \partial i} \widehat{\nu}_{b\rightarrow i}(x_i).
\end{align*}
We define \textit{the normalised information entropy of the BP fixed point}:
\begin{align*}
\beta_1=-\frac{1}{n \ln d}\sum_{i=1}^{n}\sum_{x_i=1}^{d}{\nu_{i}(x_i) \ln(\nu_{i}(x_i))}.
\end{align*}
Assign variable $i, i=1,...,n,$ the value $x_i$ that has the biggest marginal $\nu_{i}(x_i)$ among $x_i=1,...,d$. We get a BP guided assignment. We define \textit{the bias intensity of BP fixed point to the planted solution}:
$$\beta_2=\frac{\text{the overlap between the planted solution and the BP guided assignment }}{n}.$$
Bethe free entropy \cite{oxford} is $S_{Bethe}=\sum_aS_a+\sum_iS_i-\sum_{(i,a)}S_{ia}$, where$S_a=\log\left[\sum_{\underline{x}_{\partial a}}\psi_a(\underline{x}_{\partial a})\prod_{i\in \partial a} \nu_{i\rightarrow a}(x_i)\right],
S_i=\log\left[\sum_{x_i}\prod_{b\in \partial i} \widehat{\nu}_{b\rightarrow i}(x_i)\right],
S_{ia}=\log\left[\sum_{x_i}\nu_{i\rightarrow a}(x_i)\widehat{\nu}_{a\rightarrow i}(x_i)\right].
$ We define \textit{the normalised Bethe free entropy}:
$$\beta_3=\frac{S_{Bethe}}{n \ln d}.$$
\begin{figure}[htbp]
  \centering
  \includegraphics[width=0.8\columnwidth]{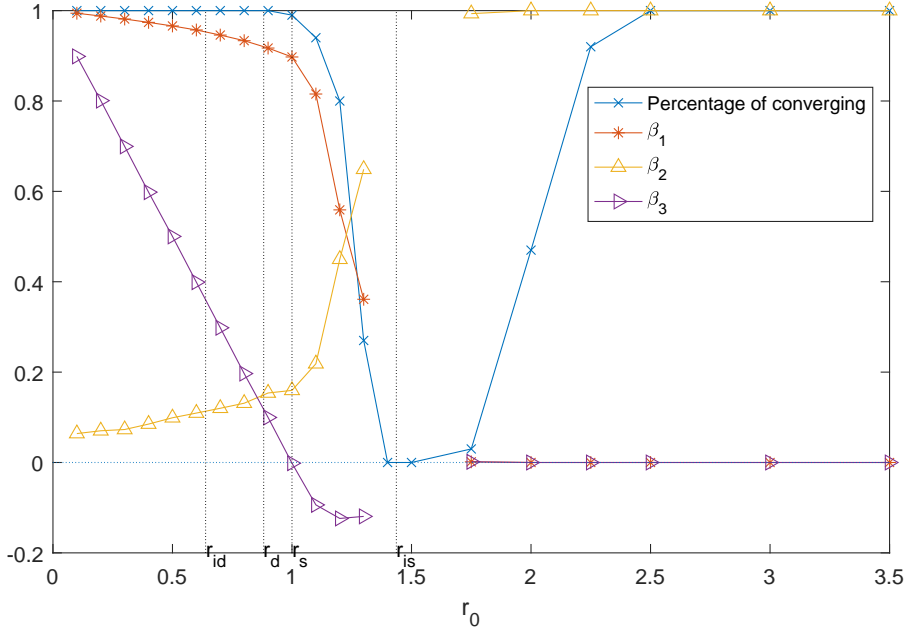}
  \caption{BP equations initialized randomly. Percentage of converging in 1000 steps over 100 planted RB instances ($n=50, k=2, \alpha=0.8, p=0.4$) is shown in the blue line. Other lines: $\beta_1$ is the normalised information entropy of the BP fixed point; $\beta_2$ is the bias intensity of BP fixed point to the planted solution; $\beta_3$ is the normalised Bethe free entropy. $r_{id}, r_d, r_s, r_{is}$ are the phase transition points. }\label{fix1}
\end{figure}

Firstly, we study the fixed point of BP equations initialized randomly, as shown in FIG. \ref{fix1}. We find that:

1. When $r<r_s$, the fixed point is liquid, which means that $\nu_{i}(x_i)\approx1/d$ for almost all $i$ and $x_i$. If $\nu_{i}(x_i)=1/d$ strictly for all $i$ and $x_i$, we have $\beta_1=1$, $\beta_2=1/23$, $\beta_3=1-r_0$. Before $r_s$ these three equations are almost established, so we say that the fixed point is liquid.

This is consistent with our results on the clustering phase, and support that the clustering phase ends at $r_s$. In the clustering phase, there are exponentially many clusters, and each cluster contains only an exponentially small proportion of solutions. Clusters are widely distributed all around the solution space, so if the BP equations give the correct marginals, the fixed point should be liquid.

2. When $r_0>2.5$, the fixed point is stable and totally biased to the planted solution. In this region $\beta_1=0$ and $\beta_2=0$. It means that, for variable $i (i=1,...,n)$, $\nu_{i}(x_i)=1$ if $x_i$ is the value of $i$ in $S$, and $\nu_{i}(x_i)=0$ otherwise. In this region the planted solution is the only solution, so in this region marginals from BP equations are correct.

3. When $1<r_0<2.5$, the BP iteration does not always converge, and for $r_0=1.1, 1.2, 1.3$ the marginals are not correct. According to the phase diagram, after $r_s$ only the small planted cluster exists, so the marginals should be biased to the planted solution strongly. But $\beta_1$ and $\beta_2$ change slowly after $r_s$, so at $r_0=1.1, 1.2, 1.3$ the marginals are not correct.

We can see that BP does not always give correct marginals. This should be related to the solution space structure. In the $r<r_s$ situation, the solutions distribute widely all around the solution space; in the over constraint situation, the configuration space has only a single energy valley. Those should be two typical situations where the BP equations give correct marginals. If the solutions do not distribute uniformly, and the configuration space has many energy valleys, the BP may fail.

\begin{figure}[htbp]
  \centering
  \includegraphics[width=0.8\columnwidth]{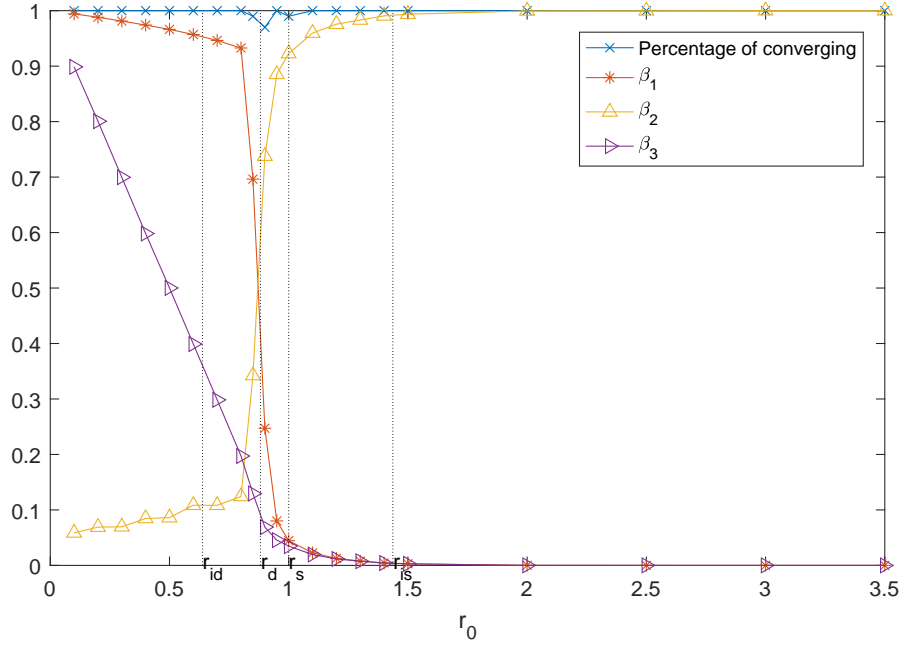}
  \caption{BP equations initialized in the planted solution. Percentage of converging in 1000 steps over 100 planted RB instances ($n=50, k=2, \alpha=0.8, p=0.4$) is shown in the blue line. Other lines: $\beta_1$ is the normalised information entropy of the BP fixed point; $\beta_2$ is the bias intensity of BP fixed point to the planted solution; $\beta_3$ is the normalised Bethe free entropy. $r_{id}, r_d, r_s, r_{is}$ are the phase transition points. }\label{fix2}
\end{figure}

Secondly, we study the fixed point of BP equations initialized in the planted solution $S$, as shown in FIG. \ref{fix2}. We find that:

1. When $r<r_d$, BP converges to the liquid solution. This is similar to the BP initialized randomly.

2. When $r=r_d$, the values of $\beta_1$ and $\beta_2$ change drasticly, and the fixed point becomes biased to $S$ suddenly.

According to [18], if BP equations is initialized in the planted solution then the fixed point is biased towards the planted solution above the reconstruction threshold $r_d$, so FIG. \ref{fix2} supports our calculation of $r_d$.

3. After the dramatic change at $r=r_d$, the values of $\beta_1, \beta_2, \beta_3$ change slowly, and they do not notice that a transition happens at $r_s$.

We can see that when $r<r_d$, the fixed point gives marginals of all solutions, while when $r>r_d$ the fixed point gives marginals of the planted cluster. With $r$ growing, the planted cluster shrinks, so the fixed point is biased to $S$ more and more.

4. At $r=r_{is}$, we have $\beta_1=0, \beta_2=1, \beta_3=0$. The fixed point is biased to $S$ totally.

This is because the planted cluster shrinks to a single solution at $r_{is}$. So FIG. \ref{fix2} supports our calculation of $r_{is}$.

\begin{figure}[htbp]
  \centering
  \includegraphics[width=0.8\columnwidth]{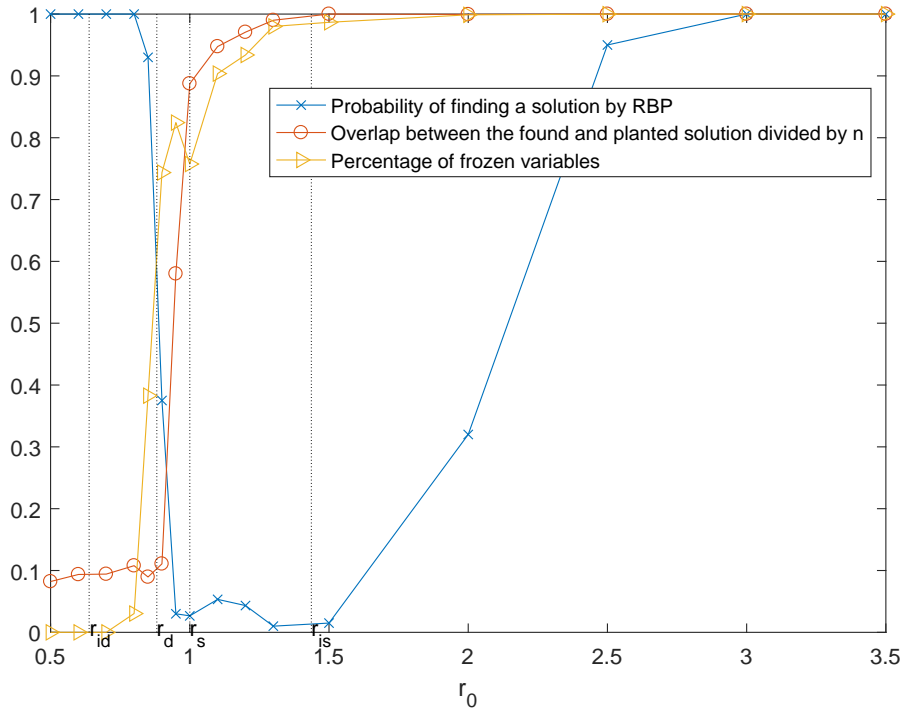}
  \caption{RBP on the planted RB model. Probability of finding a solution by RBP in 1000 steps over 300 planted RB instances ($n=50, k=2, \alpha=0.8, p=0.4$) is shown in the blue line. Overlap between the found and the planted solution per variable and percentage of frozen variables in the found solution are in the red and the orange lines. $r_{id}, r_d, r_s, r_{is}$ are the phase transition points. }\label{rbp}
\end{figure}

Thirdly, we study the solutions found by the reinforced belief propagation algorithm (RBP, referring to \cite{rbp}).
RBP algorithm finds solutions by iterations, and at each step it follows update rules (\ref{ite2}), (\ref{ite1}) with probability $p$, and update rules (\ref{ite2}), (\ref{ite3}), (\ref{ite4}) with probability $1-p$. At step $t$, we set $p=t^{-0.1}$. Update rules are
\begin{align}
\widehat{\nu}_{a\rightarrow i}^{(t)}(x_i)\cong\sum_{\underline{x}_{\partial a\setminus i}}\psi_a(\underline{x}_{\partial a})\prod_{k\in \partial a\setminus i} \nu_{k\rightarrow a}^{(t)}(x_k),\label{ite2}\\
\nu_{i\rightarrow a}^{(t+1)}(x_i)\cong\prod_{b\in \partial i\setminus a} \widehat{\nu}_{b\rightarrow i}^{(t)}(x_i),\label{ite1}\\
\nu_{i\rightarrow a}^{(t+1)}(x_i)\cong\mu_{i}^{(t)}(x_i)\prod_{b\in \partial i\setminus a} \widehat{\nu}_{b\rightarrow i}^{(t)}(x_i),\label{ite3}\\
\mu_{i}^{(t)}(x_i)\cong\prod_{b\in \partial i} \widehat{\nu}_{b\rightarrow i}^{(t-1)}(x_i).\label{ite4}
\end{align}
In each step, the algorithm estimates marginals by
\begin{align*}
\nu_{i}(x_i)\cong\prod_{b\in \partial i} \widehat{\nu}^{(t)}_{b\rightarrow i}(x_i),
\end{align*}
and verifies whether the marginals guided assignment is a solution.
In FIG. \ref{rbp}, we show the probability of finding a solution by RBP, the overlap between the found solution and the planted solution per variable, and percentage of frozen variables of the found solution. The percentage of frozen variables is obtained by applying the whitening procedure \cite{whi}.

From FIG. \ref{rbp} we can find that:

1. At $r=r_d$ the frozen variables increases rapidly.

According to the phase diagram, at $r=r_d$, solutions begin to shatter into clusters, and each cluster contains only an exponentially small proportion of solutions, so at $r=r_d$ the solutions become frozen rapidly. FIG. \ref{rbp} supports the location of $r_d$.

2. From $r=r_s$ to $r=r_{is}$, both the percentage of frozen variables and the overlap increase from about 0.8,0.9 to 1 gradually.

This is because when $r>r_s$ only the planted cluster is left, so the found solution must belong to the planted cluster. With the planted cluster shrinking, the found solution becomes totally frozen and equals to $S$ at $r_{is}$. FIG. \ref{rbp} supports the locations of $r_s$ and $r_{is}$.

Although the experiments confirm the locations of $r_d$, $r_s$ and $r_{is}$, and support the solution space structure we have gotten, we have to say that $n=50$ is not the case in the thermodynamic limit, so the experiment results do not consist with the phase diagram strictly. For example, according to the phase diagram, the planted cluster is very small, so the overlap in FIG. \ref{rbp} should be closer to 1 at $r_0=1.1, 1.2, 1.3$; the $\beta_1, \beta_2$ in FIG. \ref{fix1}, FIG. \ref{fix2}, should change more suddenly at transition points.

\section{Conclusion}
We find four phase transitions of the planted RB model. When the independent phase transition (at $r=r_{id}$) happens, the planted cluster is separated from the majority of solutions. From then on, the majority and the planted cluster have different developments: the majority of solutions will go through the clustering transition (at $r=r_d$) and the satisfiability transition (at $r=r_s$); the planted cluster will go through the isolated transition (at $r=r_{is}$). $r_{is}<r_d$, $r_d<r_{is}<r_s$ and $r_{is}>r_s$ are all possible, depending on parameters $\alpha, p, k$. The locations of $r_d, r_{is}, r_s$ is supported by some experiments.

Before $r_s$, the planted and the random RB model have the same clustering phase, which means the problem hardness is similar. After $r_s$, only the planted cluster exists in the solution space, and the size of the planted cluster is very small. So, at $r=r_s$, the problem is very hard to solve. The phase diagram guarantees that this model can generate very hard satisfiable CSP benchmarks.
If $r_s>r_{is}$, the problem should be even harder (at $r=r_s$), because the planted cluster has shrunk to a point at $r=r_s$.

If $r$ is very large, the configuration space has only a single energy valley, which makes the instances tractable for local search algorithms. With $r$ growing, a easy/hard/easy pattern is observed. The hard/easy transition should be further studied.

\section{Acknowledgments}
Project supported by the National Natural Science Foundation of China (No. 11801028).

\section{appendix A}
In appendix A we will prove (\ref{eq:fx1}). First of all, from the definition of the planted RB model we will give the expression of $\mathbb{E}(Z(x))$ in the following formula (\ref{eq:ez}). Secondly, the normalized version $f(x)$ will be given in (\ref{eq:fx2}). Thirdly the operation of finding the maximum value in (\ref{eq:fx2}) will be carried out and then (\ref{eq:fx1}) will be proven.

Let $A$ and $B$ be two assignments at Hamming distance $xn$. The number of such assignment-pairs is $$\left(\begin{array}{c}n\\xn,n-xn\end{array}\right)d^n(d-1)^{xn}.$$ Because of the planted solution, not all those assignment-pairs have the same opportunities to be solution-pairs. We have to consider the distance between $A$ and $S$, and the distance between $B$ and $S$, where $S$ is the solution we planted. We denote by $x_i(A), x_i(B), x_i(S)$ the values of variable $x_i (i\in\{1 ,... , n\})$ in assignments $A$, $B$, $S$. The $n$ variables $x_1, x_2, ..., x_n$ can be divided into 5 sets:
\begin{enumerate}
\item $\{x_i|x_i(A)=x_i(S), x_i(B)\neq x_i(S), x_i(A)\neq x_i(B), i=1,...,n\}$;
\item $\{x_i|x_i(A)\neq x_i(S)$, $x_i(B)=x_i(S)$, $x_i(A)\neq x_i(B), i=1,...,n\}$;
\item $\{x_i|x_i(A)\neq x_i(S)$, $x_i(B)\neq x_i(S)$, $x_i(A)\neq x_i(B), i=1,...,n\}$;
\item $\{x_i|x_i(A)\neq x_i(S)$, $x_i(B)\neq x_i(S)$, $x_i(A)=x_i(B), i=1,...,n\}$;
\item $\{x_i|x_i(A)=x_i(S)$, $x_i(B)=x_i(S)$, $x_i(A)=x_i(B), i=1,...,n\}.$
\end{enumerate}
Sizes of the sets are $an, bn, cn, dn, en$ respectively, then $a+b+c=x, d+e=1-x$. In such a way we classify assignment-pairs into types labeled by a tuple $(a, b, c, d, e)$. The number of assignment-pairs of type $(a, b, c, d, e), a+b+c=x, d+e=1-x,$ is $$F_1=\left(\begin{array}{c}n\\xn,n-xn\end{array}\right)\left(\begin{array}{c}xn\\an,bn,cn\end{array}\right)\left(\begin{array}{c}n-xn\\dn,en\end{array}\right)(d-1)^{(1-e)n}(d-2)^{(x-a-b)n}.$$

We will show that assignment-pairs of the same type have the same opportunity of being solution-pairs. According to the steps of generating a planted RB instances, $t=rn\ln n$ constraints are chosen randomly. Constraint $a$ involves $k$ variables which are chosen randomly, denoted by $\underline{x}_a$. Let the values of those $k$ variables in assignments $A$, $B$, $S$ be $\underline{x}_a(A)$, $\underline{x}_a(B)$, $\underline{x}_a(S)$ respectively. The relations among $\underline{x}_a(A)$, $\underline{x}_a(B)$ and $\underline{x}_a(S)$ are in 3 sorts:
 \begin{enumerate}
\item $\underline{x}_a(A)=\underline{x}_a(B)=\underline{x}_a(S)$;
\item $\underline{x}_a(A)=\underline{x}_a(B)\neq\underline{x}_a(S)$ or  $\underline{x}_a(A)\neq\underline{x}_a(B)=\underline{x}_a(S)$ or  $\underline{x}_a(B)\neq\underline{x}_a(A)=\underline{x}_a(S)$;
\item $\underline{x}_a(A)\neq\underline{x}_a(B)$ and $\underline{x}_a(A)\neq\underline{x}_a(S)$ and $\underline{x}_a(B)\neq\underline{x}_a(S)$.
\end{enumerate}
For constraint $a$, $(1-p)d^k$ compatible tuples of values are selected, including $\underline{x}_a(S)$.
For the first sort, because $\underline{x}_a(S)$ satisfies the constraint $a$, so $\underline{x}_a(A)$ and $\underline{x}_a(B)$ both satisfy $a$. The probability of the occurrence of the first sort is the probability that the randomly chosen $k$ variables all belong to the above fifth set (whose size is $en$), and the probability is the following $s_0$. For the second and the third sorts, the probabilities of occurrences are the following $s_1, s_2$; the probabilities that both $\underline{x}_a(A)$ and $\underline{x}_a(B)$ satisfy the constraint $a$ are the following $p_1$, $p_2$.
The opportunity that a assignment-pair of type $(a, b, c, d, e)$ is a solution-pair is $$F_2=(s_0+s_1p_1+s_2p_2)^t,$$ where
\begin{eqnarray*}
&&s_0=\left.\left(\begin{array}{c}en\\k,en-k\end{array}\right)\middle/\left(\begin{array}{c}n\\k,n-k\end{array}\right)\right.,\\
&&s_1=s_3+s_4+s_5-3s_0,\\
&&s_2=1-s_3-s_4-s_5+2s_0,\\
&&p_1=\left.\left(\begin{array}{c}d^k-2\\(1-p)d^k-2,pd^k\end{array}\right)\middle/\left(\begin{array}{c}d^k-1\\(1-p)d^k-1,pd^k\end{array}\right)\right.,\\
&&p_2=\left.\left(\begin{array}{c}d^k-3\\(1-p)d^k-3,pd^k\end{array}\right)\middle/\left(\begin{array}{c}d^k-1\\(1-p)d^k-1,pd^k\end{array}\right)\right.,
\end{eqnarray*}
and
\begin{eqnarray*}
&&s_3=\left.\left(\begin{array}{c}(a+e)n\\k,(a+e)n-k\end{array}\right)\middle/\left(\begin{array}{c}n\\k,n-k\end{array}\right)\right.,\\
&&s_4=\left.\left(\begin{array}{c}(b+e)n\\k,(b+e)n-k\end{array}\right)\middle/\left(\begin{array}{c}n\\k,n-k\end{array}\right)\right.,\\
&&s_5=\left.\left(\begin{array}{c}n-xn\\k,n-xn-k\end{array}\right)\middle/\left(\begin{array}{c}n\\k,n-k\end{array}\right)\right..
\end{eqnarray*}
The expectation of the number of solutions pairs at distance $xn$ is
\begin{align}\label{eq:ez}
\mathbb{E}(Z(x))=\sum_{\begin{array}{c}an+bn+cn=xn,\\dn+en=n-xn\end{array}}{ F_1\cdot F_2}
\end{align}

This summation operation contains a polynomial number of items, so $\lim_{n \rightarrow \infty}\frac{\ln \left(\mathbb{E}(Z(x))\right)}{n\ln n}$ is determined by the largest term.
With $n\rightarrow \infty$, we have $p_1\rightarrow 1-p, p_2\rightarrow (1-p)^2, s_0\rightarrow e^k, s_3\rightarrow (a+e)^k, s_4\rightarrow (b+e)^k, s_5\rightarrow (1-x)^k$, and by the Stirling formula, we have
\begin{align*}
\frac{\ln{\left(\begin{array}{c}n\\xn,n-xn\end{array}\right)}}{n}\rightarrow -\ln(x^x(1-x)^{1-x}).
\end{align*}
The formula $f(x)$ defined in (\ref{eq:fx0}) can be simplified:
\begin{eqnarray}
f(x)&&=\lim_{n \rightarrow \infty}\frac{\ln \left(\mathbb{E}(Z(x))\right)}{n\ln n}\nonumber\\
&&=\lim_{n \rightarrow \infty}\frac{\ln \left(\max_{\substack{an+bn+cn=xn,\\dn+en=n-xn}}F_1\cdot F_2\right)}{n\ln n}\nonumber\\
&&=\max_{\substack{0\leq a,b \leq x,\\0\leq a+b \leq x, \\0\leq e \leq 1-x}}[\alpha(1+x-e-a-b)+r\ln A(x,a,b,e)],\label{eq:fx2}
\end{eqnarray}
where
\begin{eqnarray*}
&&A(x,a,b,e)\\
=&&e^k+\left[(1-x)^k+(a+e)^k+(b+e)^k-3e^k\right](1-p)+\left[1-(1-x)^k-(a+e)^k-(b+e)^k+2e^k\right](1-p)^2\\
=&&(1-p)^2+\left[1-3(1-p)+2(1-p)^2\right]e^k+\left[(1-p)-(1-p)^2\right]\left[(1-x)^k+(a+e)^k+(b+e)^k\right].
\end{eqnarray*}

Let $a+b=m$. $(a+e)^k+(b+e)^k$ reaches maximum when $a=0, b=m$, so
\begin{eqnarray}\label{eq:lll}
&&\max_{\substack{0\leq a,b \leq x,\\0\leq a+b \leq x, \\0\leq e \leq 1-x}} A(x,a,b,e)=\max_{\substack{0\leq m \leq x, \\0\leq e \leq 1-x}}(1-p)^2+p^2e^k+[(1-p)-(1-p)^2]\left((1-x)^k+(m+e)^k\right)
\end{eqnarray}
Let $m+e=l$. If $l<1-x$, the above formula reach maximum at $e=l$; if $l>1-x$, the above formula reach maximum at $e=1-x$; then
$$(\ref{eq:lll})=\max\{A_1, A_2\},$$ where
\begin{eqnarray*}
A_1=\max_{0\leq l\leq1-x}r\ln[(1-p)^2+(1-p)p(1-x)^k+pl^k],\\
A_2=\max_{1-x<l\leq 1}r\ln[(1-p)^2+p(1-x)^k+(1-p)pl^k].
\end{eqnarray*}
Combining (\ref{eq:fx2}) and (\ref{eq:lll}), we have $$f(x)=\max\{\max_{0\leq l\leq1-x}f_3, \max_{1-x<l\leq 1}f_4\},$$ where
\begin{eqnarray*}
f_3=\alpha(1+x-l)+r\ln[(1-p)^2+(1-p)p(1-x)^k+pl^k], \\
f_4=\alpha(1+x-l)+r\ln[(1-p)^2+p(1-x)^k+(1-p)pl^k].
\end{eqnarray*}
Calculate the second order differentials of $f_3(x,l)$ and $f_4(x,l)$ to $l$,
\begin{eqnarray*}
\frac{\partial^2 f_3}{\partial l^2}=\frac{rpkl^{k-2}\left[(k-1)\left((1-p)^2+(1-p)p(1-x)^k+pl^k\right)-pkl^{k}\right]}{[(1-p)^2+(1-p)p(1-x)^k+pl^k]^2}, \\
\frac{\partial^2 f_4}{\partial l^2}=\frac{r(1-p)pkl^{k-2}\left[(k-1)\left((1-p)^2+p(1-x)^k+(1-p)pl^k\right)-(1-p)pkl^{k}\right]}{[(1-p)^2+p(1-x)^k+(1-p)pl^k]^2},
\end{eqnarray*}
then we find $\frac{\partial^2 f_3}{\partial l^2}\geq 0$ $(0\leq l\leq1-x)$, $\frac{\partial^2 f_4}{\partial l^2}\geq 0$ $(1-x<l\leq 1)$ under the condition $p<1-\frac{1}{k}$.
Then $\max_{0\leq l\leq1-x}f_3$ and $\max_{1-x<l\leq 1}f_4$ can only be achieved on the endpoints, i.e.
$$f(x)=\max\{f_3(l=0), f_3(l=1-x), f_4(l=1-x), f_4(l=1)\},$$ and further calculating shows $$f(x)=\max\{f_1(x), f_2(x)\},$$ where $f_1(x)$ and $f_2(x)$ are defined in (\ref{eq:f1}) (\ref{eq:f2}).


\begin{thebibliography}{}
\bibitem{oxford}{M{\'{e}}zard M, Montanari A (2009) Information, physics, and computation. Oxford University Press.}
\bibitem{inference}{L Zdeborov\'{a},  Krzakala F . Statistical physics of inference: Thresholds and algorithms[J]. Advances in Physics, 2018, 65(5):453-552.}

\bibitem{handbook}{Rossi F, Beek P V, Walsh T. Handbook of Constraint Programming[M]// Handbook of constraint programming /. Elsevier, 2006.}

\bibitem{Gent}{Gent I P, Macintyre E, Prosser P, et al. Random Constraint Satisfaction: Flaws and Structure[J]. Constraints, 2001, 6(4):345-372.}

\bibitem{Smith}{Smith B M, Dyer M E. Locating the phase transition in binary constraint satisfaction problems[J]. Artificial Intelligence, 1996, 81(s 1-2):155-181.}

\bibitem{ach97} Achlioptas D, Kirousis L M, Kranakis E, et al. Random constraint satisfaction: A more accurate picture[C]// International Conference on Principles and Practice of Constraint Programming. Springer Berlin Heidelberg, 1997:107-120.
\bibitem{Lecoutre}{Lecoutre C. Constraint Networks: Techniques and Algorithms. john wiley \& sons, 2009.}
\bibitem{xu2000} Xu K, Li W. Exact Phase Transitions in Random Constraint Satisfaction Problems.[J]. Journal of Artificial Intelligence Research, 2000, 12(1):93-103.
\bibitem{bar2002}{Barthel W, Hartmann A K, Leone M, et al. Hiding solutions in random satisfiability problems: a statistical mechanics approach[J]. Physical Review Letters, 2002, 88(18):188701.}
\bibitem{jia2004}{Jia H, Moore C, Selman B. From Spin Glasses to Hard Satisfiable Formulas[C]// International Conference on Theory and Applications of Satisfiability Testing. Springer Berlin Heidelberg, 2004:199-210.}
\bibitem{locked}{L Zdeborov\'{a},  Krzakala F . Quiet Planting in the Locked Constraint Satisfaction Problems[J]. SIAM Journal on Discrete Mathematics, 2009, 25(2):750-770.}
\bibitem{ke}{Xu K, Boussemart F, Hemery F ,et al. Random constraint satisfaction: easy generation of hard (satisfiable) instances[J]. Artificial Intelligence, 2007, 171:514-534.}
\bibitem{web}{http://www.nlsde.buaa.edu.cn/~kexu/benchmarks/graph-benchmarks.htm.}
\bibitem{cai}{Cai S, Su K and Sattar A 2011 Local search with edge weighting and configuration checking heuristics for minimum vertex cover Artif. Intell. 175 1672-96.}
\bibitem{rosin}{Rosin C D . Unweighted Stochastic Local Search can be Effective for Random CSP Benchmarks[J]. Computer Science, 2014.}

\bibitem{pair1}{Achlioptas D ,  Coja-Oghlan A ,  F  Ricci-Tersenghi. On the Solution-Space Geometry of Random Constraint Satisfaction Problems[J]. Random Structures \& Algorithms, 2011, 38.}
\bibitem{xuwei}{Xu W, Zhang P, Liu T, et al. Solution space structure of random constraint satisfaction problems with growing domains[J]. Journal of Statistical Mechanics Theory \& Experiment, 2015, 2015(12):P12006.}

\bibitem{pair2}{Achlioptas D . Solution clustering in random satisfiability[J]. European Physical Journal B, 2008, 64(3-4):395-402.}
\bibitem{pair3}{Mezard M ,  Mora T ,  Zecchina R . Clustering of solutions in the random satisfiability problem[J]. Physical Review Letters, 2005, 94(19):197205.}


\bibitem{xuwei2}{Xu W ,  Gong F ,  Zhou G . Clustering phase of a general constraint satisfaction problem model d - k -CSP[J]. Physica A: Statistical Mechanics and its Applications, 2019, 537:122708.}

\bibitem{hiding}{Krzakala F , L Zdeborov\'{a}. Hiding Quiet Solutions in Random Constraint Satisfaction Problems[J]. Physical Review Letters, 2009, 102(23):238701.}

\bibitem{why}{Coja-Oghlan A ,  Krivelevich M ,  Dan V . Why Almost All k-Colorable Graphs Are Easy[J]. Theory of Computing Systems, 2010, 46(3):523-565.}

\bibitem{rbp}{Braunstein A, Zecchina R. Learning by message-passing in networks of discrete synapses[J]. Physical Review Letters, 2005, 96(3):030201.}
\bibitem{whi}{Parisi G. On the survey-propagation equations in random constraint satisfiability problems[J]. Journal of Mathematical Physics, 2008, 49(12):812-695.}

\end{thebibliography}
\end{document}